\documentclass[manuscript]{aa}
\usepackage{graphicx}
\usepackage{multirow}
\usepackage{natbib}
\usepackage{color} 
\usepackage{amsmath}           
\usepackage{amsfonts} 

\newcommand{\kms}{km~s$^{-1}$}
\newcommand{\etal}{{\em et al.\ }}

\begin{document}
\title{Omnipresent long-period intensity oscillations in open coronal structures}
\author{S. Krishna Prasad\inst{1}, D. Banerjee\inst{1,2}, T. Van Doorsselaere\inst{2}, J. Singh\inst{1}}
\institute{Indian Institute of Astrophysics, Koramangala, Bangalore 560034, India. \and Centre for mathematical Plasma Astrophysics, Mathematics Department, KU Leuven, Celestijnenlaan 200B bus 2400, 3001 Heverlee, Belgium}
\offprints{S. Krishna Prasad \email{krishna@iiap.res.in}}
\date{Received ... / Accepted ...}

\begin{abstract}
{Quasi-periodic propagating disturbances in coronal structures have been interpreted as slow magneto-acoustic waves and/or periodic upflows. It is important to understand the nature of these disturbances before proceeding to apply them.}
{Here we aim to understand the nature of these disturbances from the observed properties using a three-hour imaging sequence from AIA/SDO in two different temperature channels. We also compare the characteristics with a simple wave model.}
{We searched for propagating disturbances in open-loop structures at three different locations; a fan loop-structure off-limb, an on-disk plume-like structure and the plume/interplume regions in the north pole of the sun. In each of the subfield regions chosen to cover these structures, the time series at each pixel location was subjected to wavelet analysis to find the different periodicities. We then constructed powermaps in three different period ranges, short (2~min -- 5~min), intermediate (5~min -- 12~min), and long (12~min -- 25~min), by adding the power in the individual range above the 99\% significance level. We also constructed space-time maps for the on-disk plume structure to estimate the propagation speeds in different channels.}
{We find propagating disturbances in all three structures. Powermaps indicate that the power in the long-period range is significant up to comparatively longer distances along the loop than that in the shorter periods. This nature is observed in all three structures. A detailed analysis on the on-disk plume structure gives consistently higher propagation speeds in the 193~\AA\ channel and also reveals spatial damping along the loop. The amplitude and the damping length values are lower in hotter channels, indicating their acoustic dependence.}
{These properties can be explained very well with a propagating slow-wave model. We suggest that these disturbances are more likely to be caused by propagating slow magneto-acoustic waves than by high-speed quasi-periodic upflows. We find that intensity oscillations in longer periods are omnipresent at larger heights even in active regions.}
\end{abstract}


\titlerunning{Oscillations in open structures}
\authorrunning{S. Krishna Prasad \etal}
\maketitle

\section{Introduction}
\label{S-intro}
Small-amplitude propagating intensity disturbances are often observed with imaging telescopes in open/extended loop structures close to active regions on the sun \citep{berghmans1999, 2000A&A...355L..23D, 2003A&A...404L...1K} and also in polar plumes \citep{1998ApJ...501L.217D, 2010A&A...510L...2M, 2011A&A...528L...4K}. These are identified as alternating slanted ridges in the space-time maps, which are thought to be signatures of slow magneto-acoustic waves owing to some of the observed properties such as periodicities and propagation speeds \citep[][and references therein]{2009SSRv..149...65D}. Assuming them to be slow-mode waves, their contribution to the coronal heating and the acceleration of the solar wind has been worked out by several authors \citep{2000A&A...355L..23D, 2006A&A...448..763M, 1998ApJ...501L.217D, 2007A&A...463..713O}. Propagating slow waves can also be used as a tool for coronal seismology. Some authors used these waves' properties to derive coronal parameters that are difficult to obtain from direct observations \citep{2009A&A...503L..25W}. But the recent detailed analyses using the spectral line profiles from these structures reveal periodically enhanced emission in blue wings, which is indicative of quasi-periodic upflows \citep{2010ApJ...722.1013D, 2011ApJ...727L..37T}. It is also argued that a quasi-static plasma component in the line of sight may induce the slow waves to produce similar spectral signatures \citep{2010ApJ...724L.194V}. This makes the general nature of these disturbances somewhat ambiguous. \citet{2011ApJ...737L..43N} reported a case of slow-mode waves propagating on top of a continuous outflow. They found that the wave signatures become dominant at larger heights. It is important to establish the nature of these disturbances before proceeding to evaluate their importance in heating or to other applications. To understand the nature of these disturbances, we study the distribution of power at different frequency bins for different open coronal structures. 
\section{Observations}
\label{S-observations} 
A three-hour-long imaging sequence taken by the Atmospheric Imaging Assembly \citep[AIA;][]{2012SoPh..275...17L} on-board SDO, on June 5, 2010 starting from 10:00:11 UT, was used in this analysis. We chose two different coronal channels centred at 171~\AA, and 193~\AA, with their peak temperature responses at 0.8 MK, and 1.25 MK, respectively. Images at level 1.0 were already subjected to basic corrections that involve removal of dark current, de-spiking, flat-fielding, and bad-pixel removal. These were then processed to co-align all images in different channels to a common centre, to correct the roll angles, and to rescale the images to a common plate scale using the \verb+aia_prep.pro+ routine (version 4.10), available in SolarSoftWare (SSW). The final pixel scale is $\approx$ 0.6\arcsec\ and the cadence of the data is 12~s.

Three different open-loop structures, one isolated on-disk plume-like structure, a set of fan loops extending off-limb, and plume/interplume structures above the north pole of the sun were chosen in this study. Square field sub regions of dimensions 350, 600, and 700 pixel units, respectively, were considered to cover these three target regions. In both channels, images corresponding to the on-disk region were tracked for solar rotation and then co-aligned using the intensity cross-correlation, taking the first image as reference. Images in the off-limb locations were co-aligned directly because it was not possible to correct for solar rotation.
\section{Analysis and results}
\label{S-results}
\begin{figure*}
\centering
\includegraphics[angle=90,height=7.8cm]{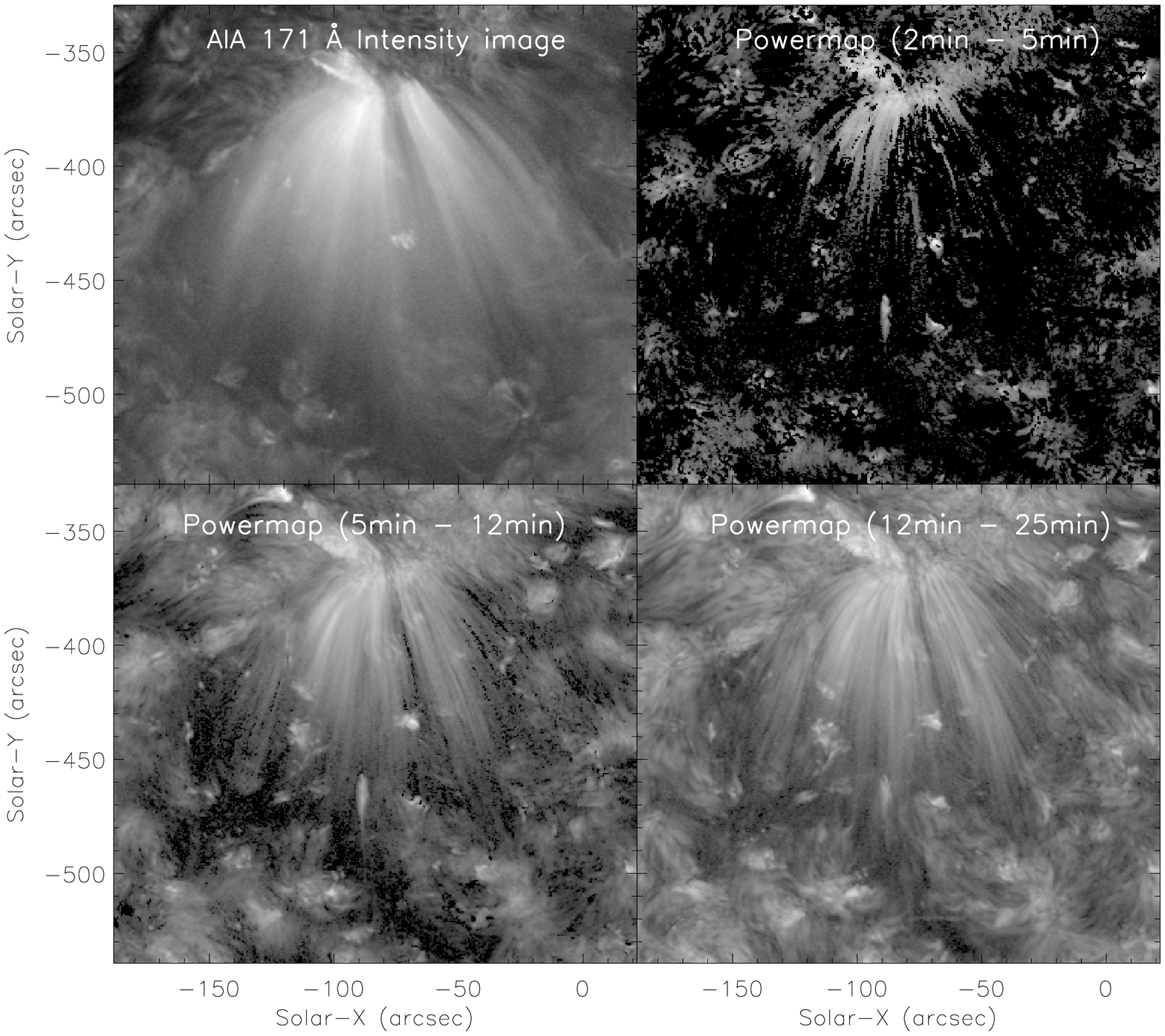}
\includegraphics[angle=90,height=7.8cm]{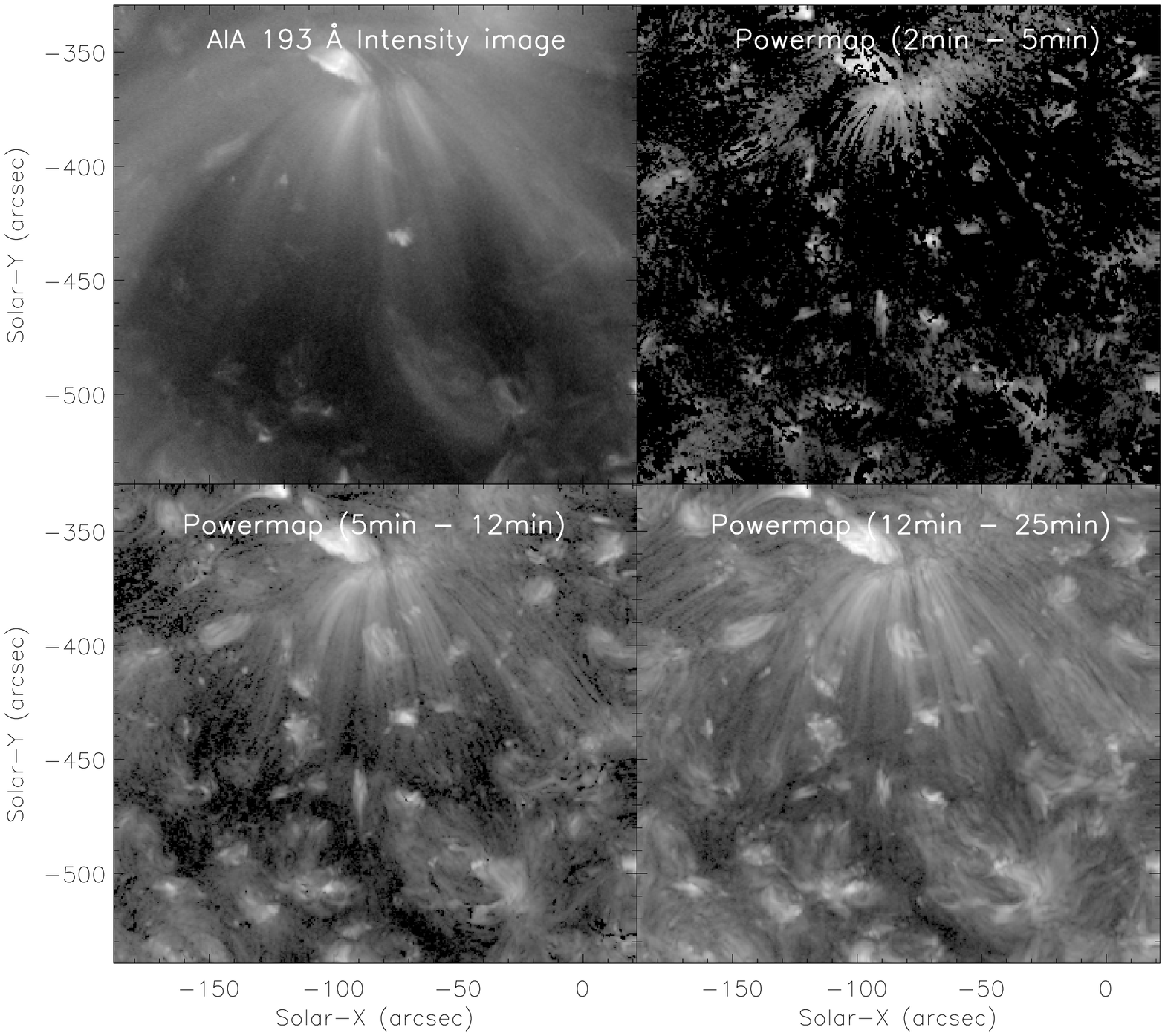}
\includegraphics[angle=90,height=7.8cm]{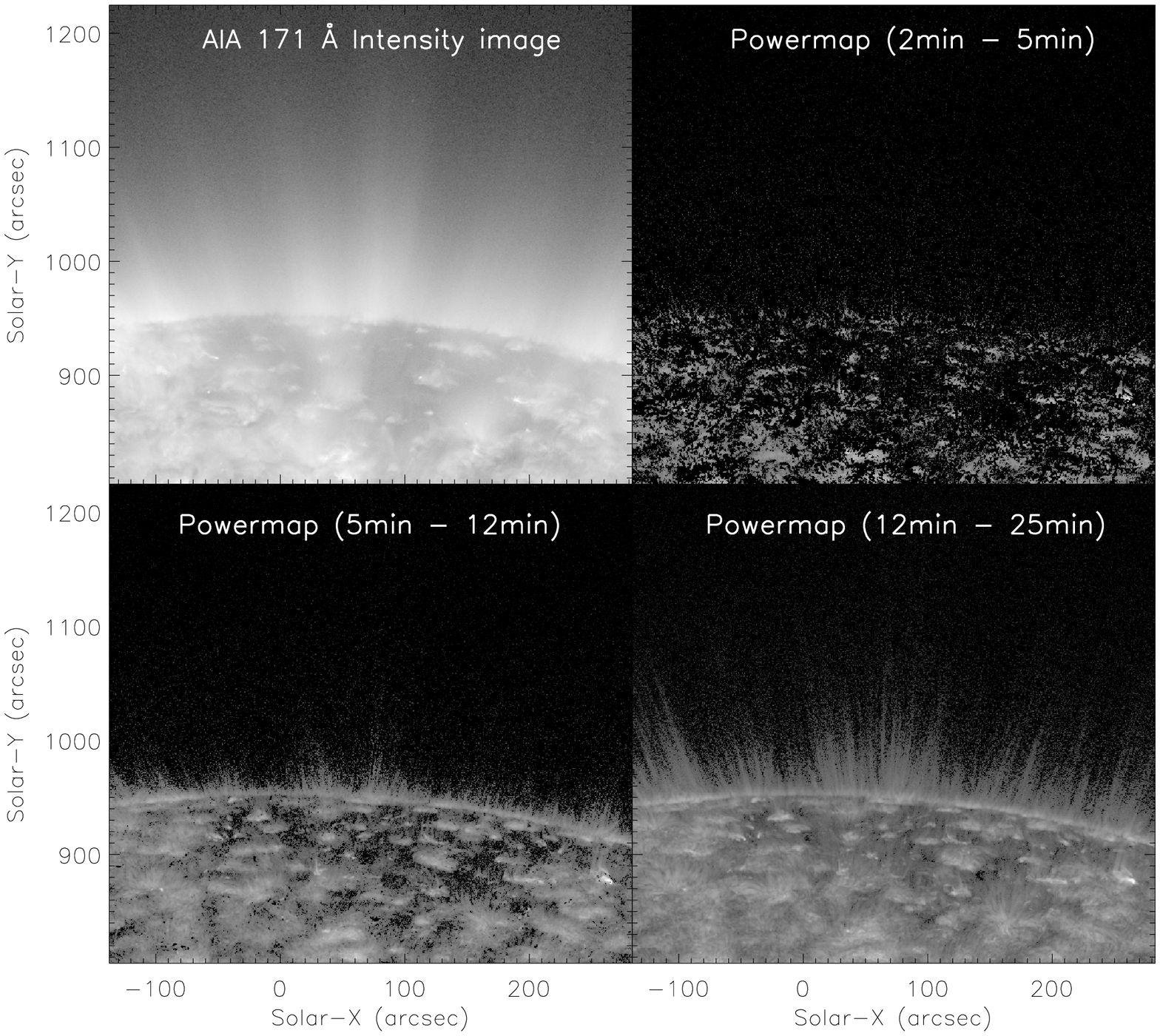}
\includegraphics[angle=90,height=7.8cm]{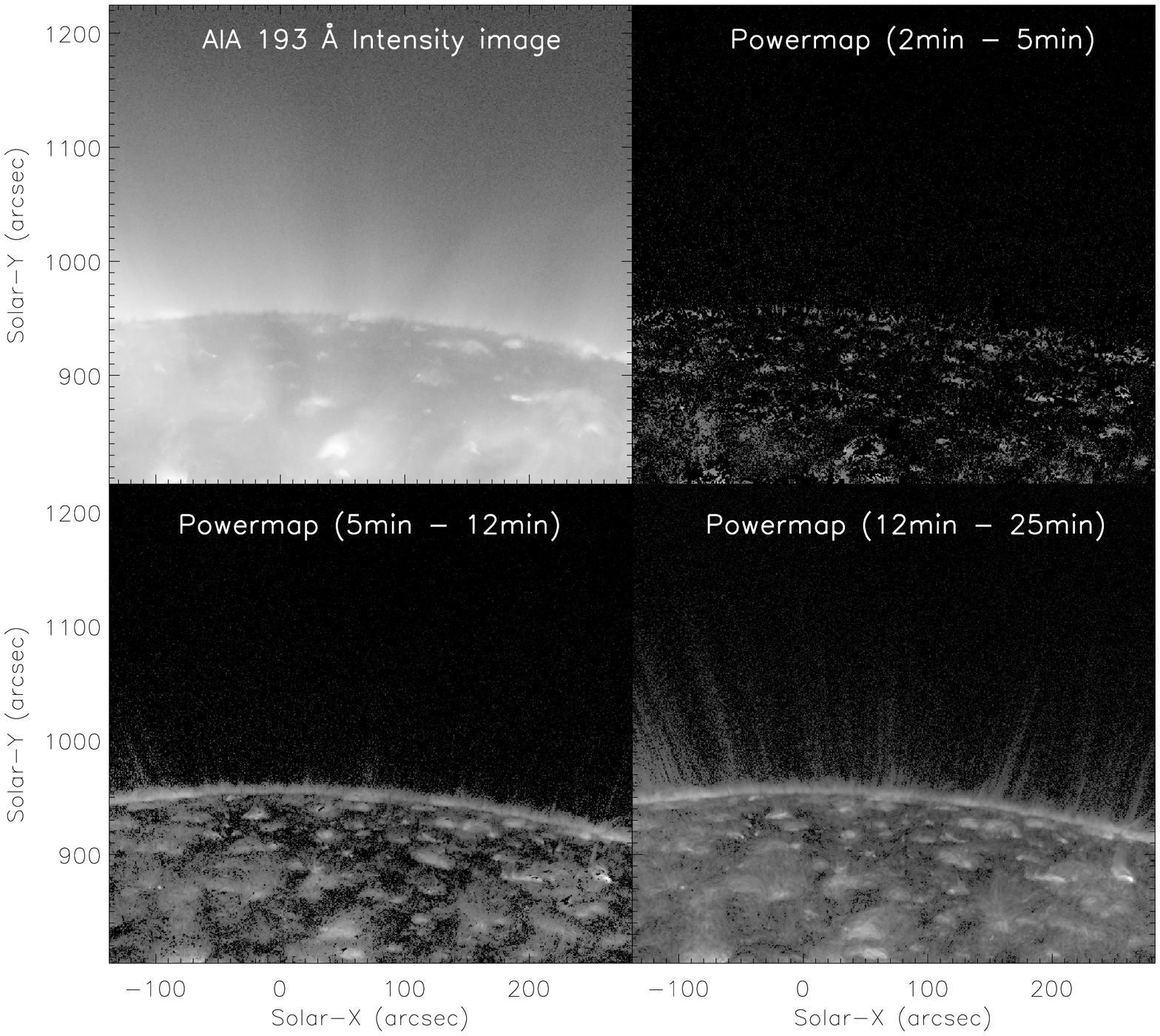}
\caption{{\it Top row:} Powermaps constructed in different periodicity ranges for the on-disk plume structure in 171~\AA\ (left) and 193~\AA\ (right) channels of AIA. The intensity image is shown in the top left panels. All powermaps show the total significant power above the 99\% confidence level in the periodicity limits. {\it Bottom row:} Same, but for the north polar region}
\label{171_193_other}
\end{figure*}

We performed a wavelet analysis for the time series at each pixel to determine the distribution of power at different frequencies. Keeping the total length of the time series (3~hrs) in mind, we decided to eliminate periods longer than 30~min. This was accomplished by detrending the time series with a 150 point ($\approx$ 30~min) smoothed original light curve before performing the wavelet analysis. Frequencies with power above the 99\% confidence level were chosen to be significant for a white noise process \citep{1998BAMS...79...61T}. We then considered three different periodicity ranges, which we refer to as short (2~min -- 5~min), intermediate (5~min -- 12~min), and long (12~min -- 25~min), and produced power maps corresponding to these by adding the total significant power within the periodicity limits at each pixel location. 
\begin{figure}
\centering
\resizebox{\hsize}{!}{\includegraphics[angle=90]{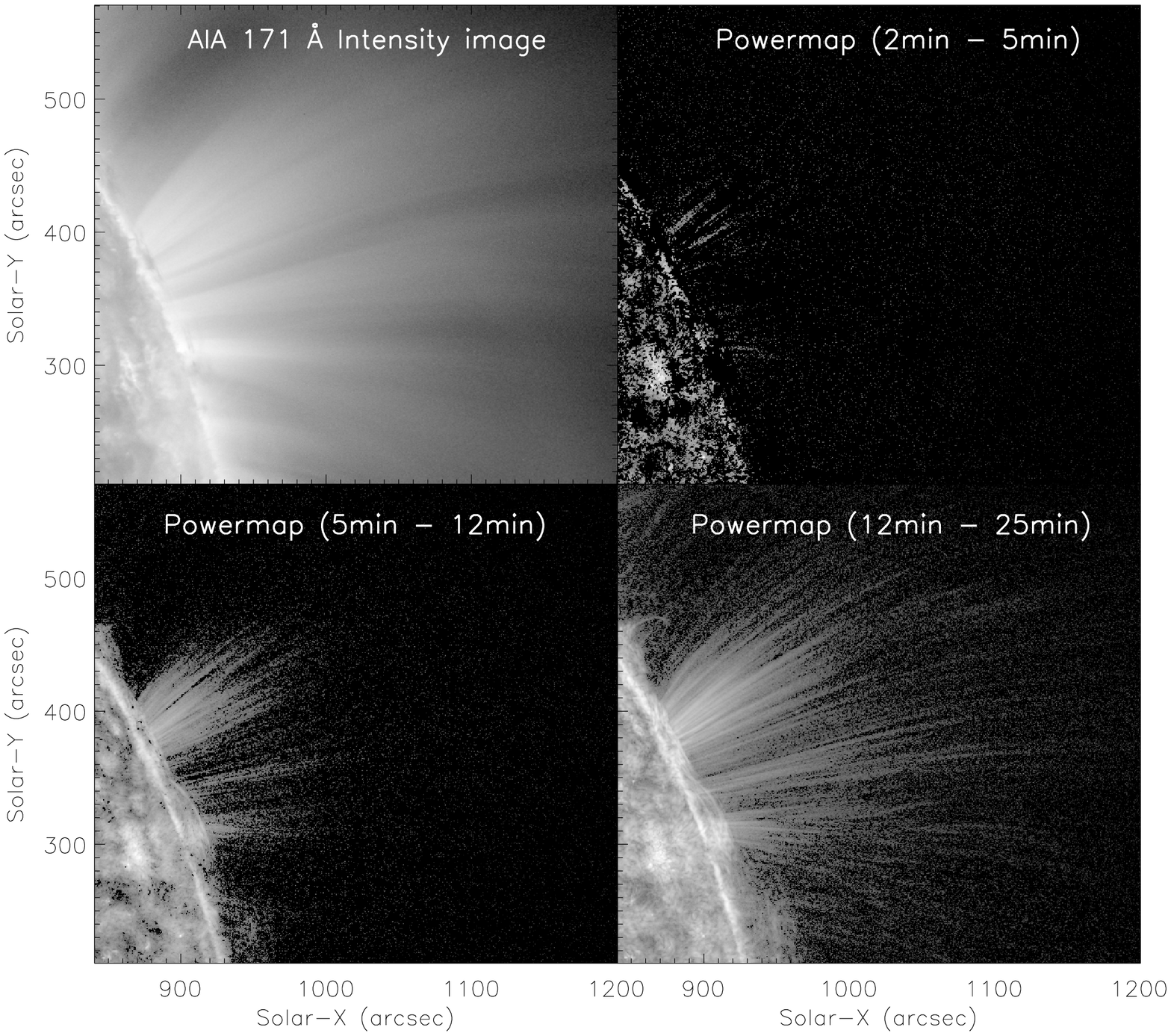}}
\resizebox{\hsize}{!}{\includegraphics[angle=90]{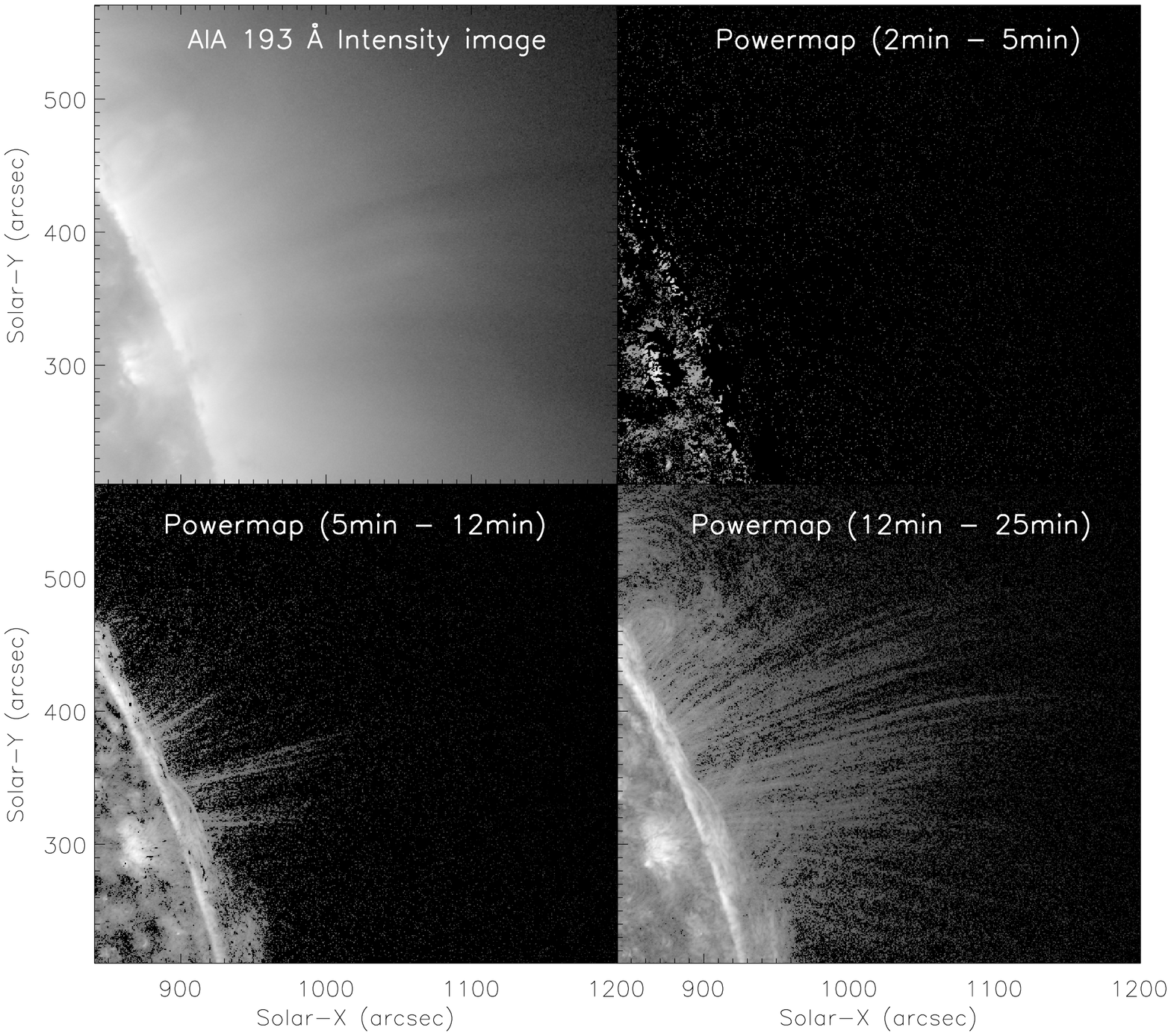}}
\caption{Powermaps in different periodicity ranges for the off-limb loop structure in the 171~\AA\ (top) and 193~\AA\ (bottom) channels of AIA. The intensity image is shown in the top left panels. All powermaps show the total power above the 99\% confidence level in the periodicity limits.}
\label{171_193_off-limb}
\end{figure}
Figure~\ref{171_193_off-limb} displays the original intensity image (top left panels) along with powermaps for these periodicity ranges in the 171~\AA\ (top) and 193~\AA\ (bottom) channels of AIA for the off-limb loop structure.  The powermaps clearly show that the power in longer periods is significant up to larger distances along the loop. In other words, this could mean that the longer periods travel farther compared to the shorter ones, which become damped faster. This trend is followed in both channels. This is true even for the other two structures, the on-disk plume-like structure and the north polar plume/interplume structures. The powermaps for these structures are shown in Fig.~\ref{171_193_other}. Since this behaviour is the same, it is possible that the nature/source of these disturbances is the same. 

We now focus on the on-disk plume structure for a more detailed analysis. We traced a portion of the loop, constructed space-time maps and processed them to enhance the visibility of the alternating ridges. 
\begin{figure*}[ht]
\centering
\begin{minipage}{0.39\textwidth}
\includegraphics[height=0.95\textwidth,angle=90]{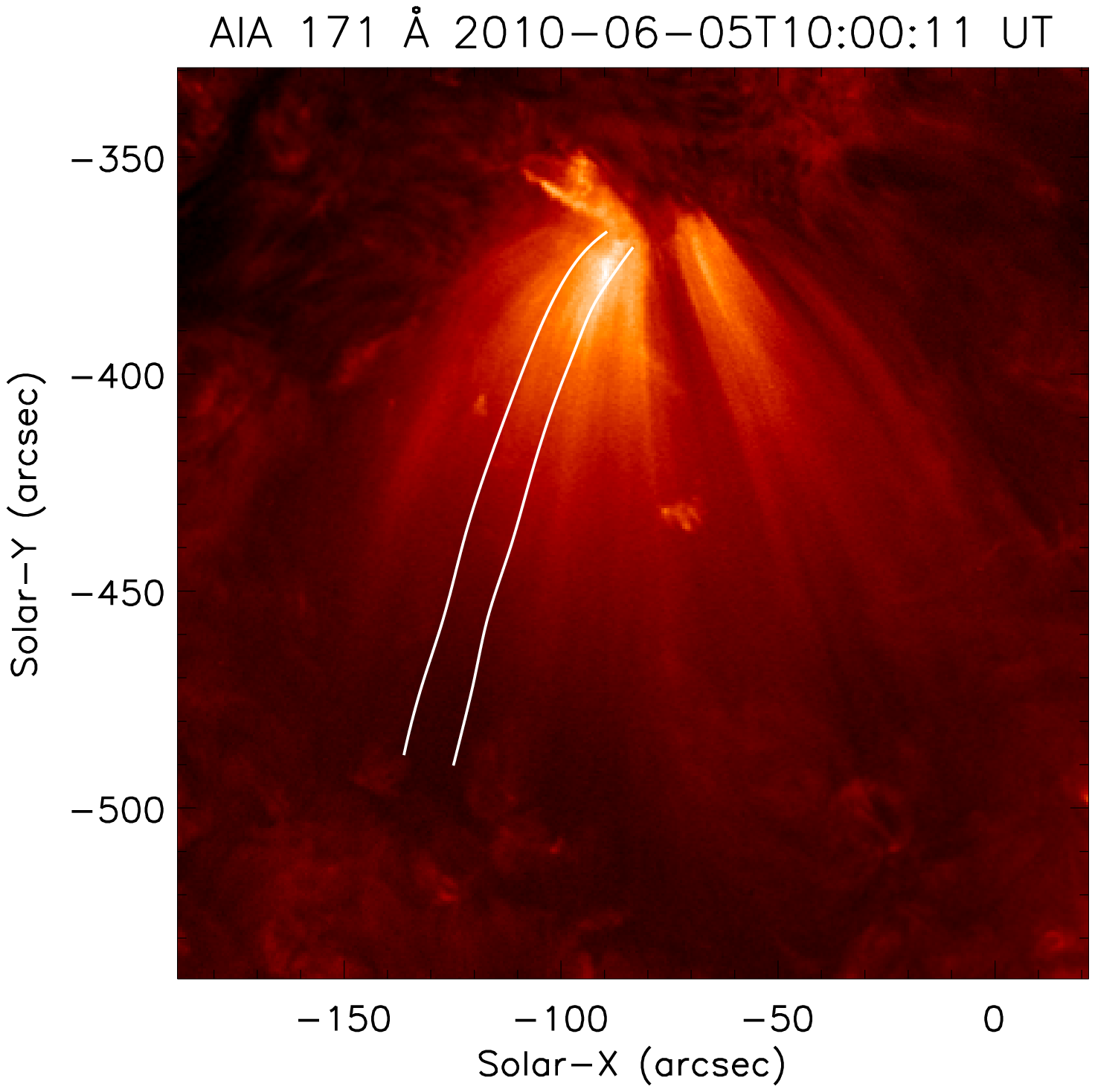}
\end{minipage}
\begin{minipage}{0.60\textwidth}
\includegraphics[height=0.98\textwidth,angle=90]{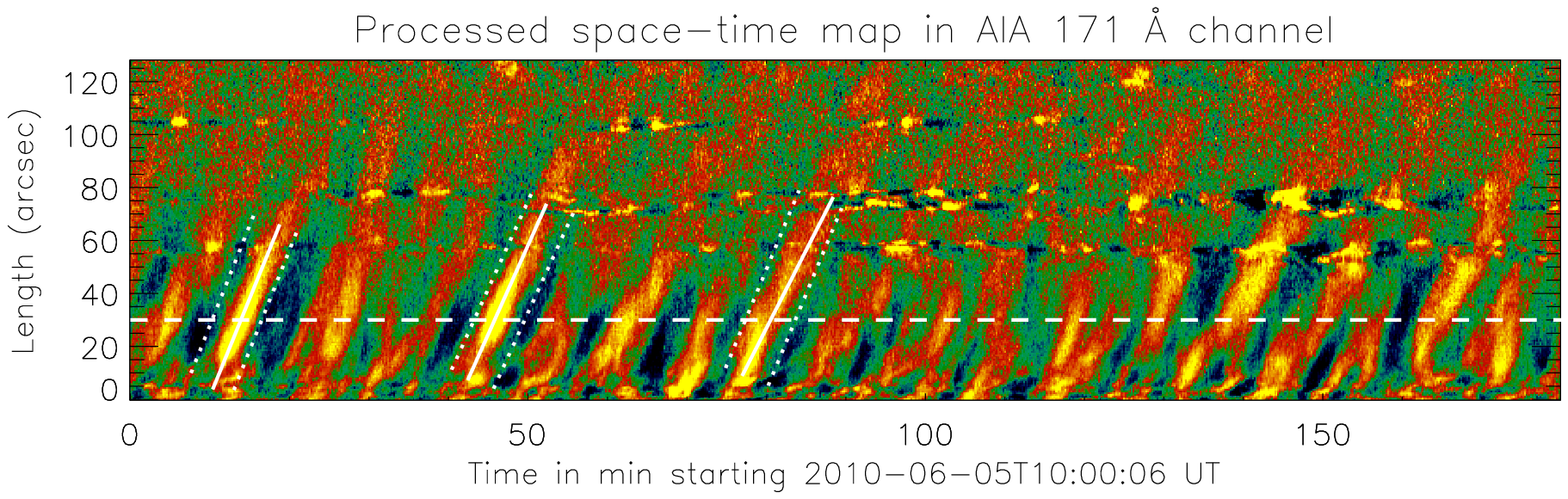}\\
\includegraphics[height=0.98\textwidth,angle=90]{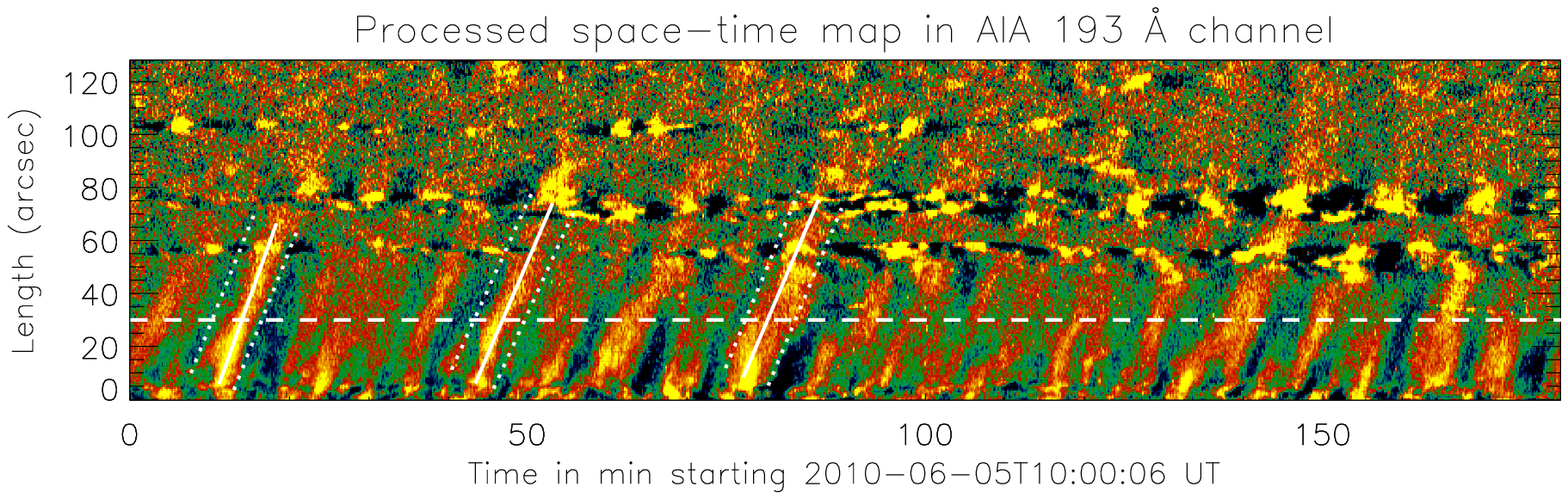}
\end{minipage}
\caption{{\it Left:} Snapshot displaying the on-disk plume-like structure. The white lines overplotted bound the chosen loop location. {\it Right:} Processed space-time maps constructed from this loop portion in the 171~\AA\ (top) and 193~\AA\ (bottom) channels. The overplotted slanted solid lines connect the final peak positions derived from the linear fit. The slopes of these lines give the propagation speeds. The horizontal dashed lines in each panel mark the location of the row used in the wavelet analysis for periodicity estimation.}
\label{on-disk_plume}
\end{figure*}
Fig.~\ref{on-disk_plume} shows the location of the loop on a snapshot displaying the plume like structure along with the processed space-time maps in the two AIA channels. The processing was performed by subtracting a background trend constructed from a 55 point ($\approx$ 11~min) smooth average of the original and then dividing by it. The trend subtraction removes or suppresses periods longer than 11~min, and division by the trend compensates for the intensity fall along the loop. We measured the propagation speed at three clean ridges following the method described in \citet{krishna2012}. This method involves identifying the peak positions in time from the space-time map at each spatial location along the ridge and then fitting them linearly. The peak positions were identified from the locations of maxima within a strip wide enough to cover the ridge alone. The apparent propagation speed and the error in its estimation could then be derived from the slope of the fit and the goodness of the fit. The boundaries of the chosen strips and the final fitted peak positions are shown as dotted and solid lines along the 
\begin{table}
\begin{center}
\caption{Periodicity and propagation speeds in different AIA channels. Second-strongest periods are also listed in brackets.}
\label{values}
\begin{tabular}{ccccc}
\hline\hline
Channel & Periodicity & \multicolumn{3}{c}{Apparent speed (\kms)} \\
\cline{3-5}
 (\AA)  &  (min)      & Ridge1 & Ridge2 & Ridge3 \\
\hline
 171    & 9.4 (15.7) & 88.7$\pm$9.7  & 80.5$\pm$5.7 & 70.6$\pm$5.8 \\
 193    & 9.4 (15.7) & 100.8$\pm$6.4 & 84.0$\pm$8.6 & 85.0$\pm$13.3 \\
\hline
\end{tabular}
\end{center}
\end{table}
ridge, respectively, in the space-time maps shown in Fig.~\ref{on-disk_plume}. Unlike the case of choosing two points manually on each ridge, this method takes several points along the ridge and uses the peak positions in the speed estimation and hence improves the error in it. Recently, \citet{2012SoPh..279..427K} used a similar method to determine the speeds of such propagating disturbances and discussed their temperature dependence. \citet{2012A&A...543A...9Y} also developed three different methods for measuring the apparent speed of propagating waves with considerably high accuracy. Apparent propagation speeds obtained from the two AIA channels are listed in Table~\ref{values}. The same ridges are chosen in both channels for comparison. The observed apparent propagation speed is consistently higher in all three ridges in the hotter 193~\AA\ channel. 
\begin{figure*}
\centering
\includegraphics[angle=90,height=8.0cm]{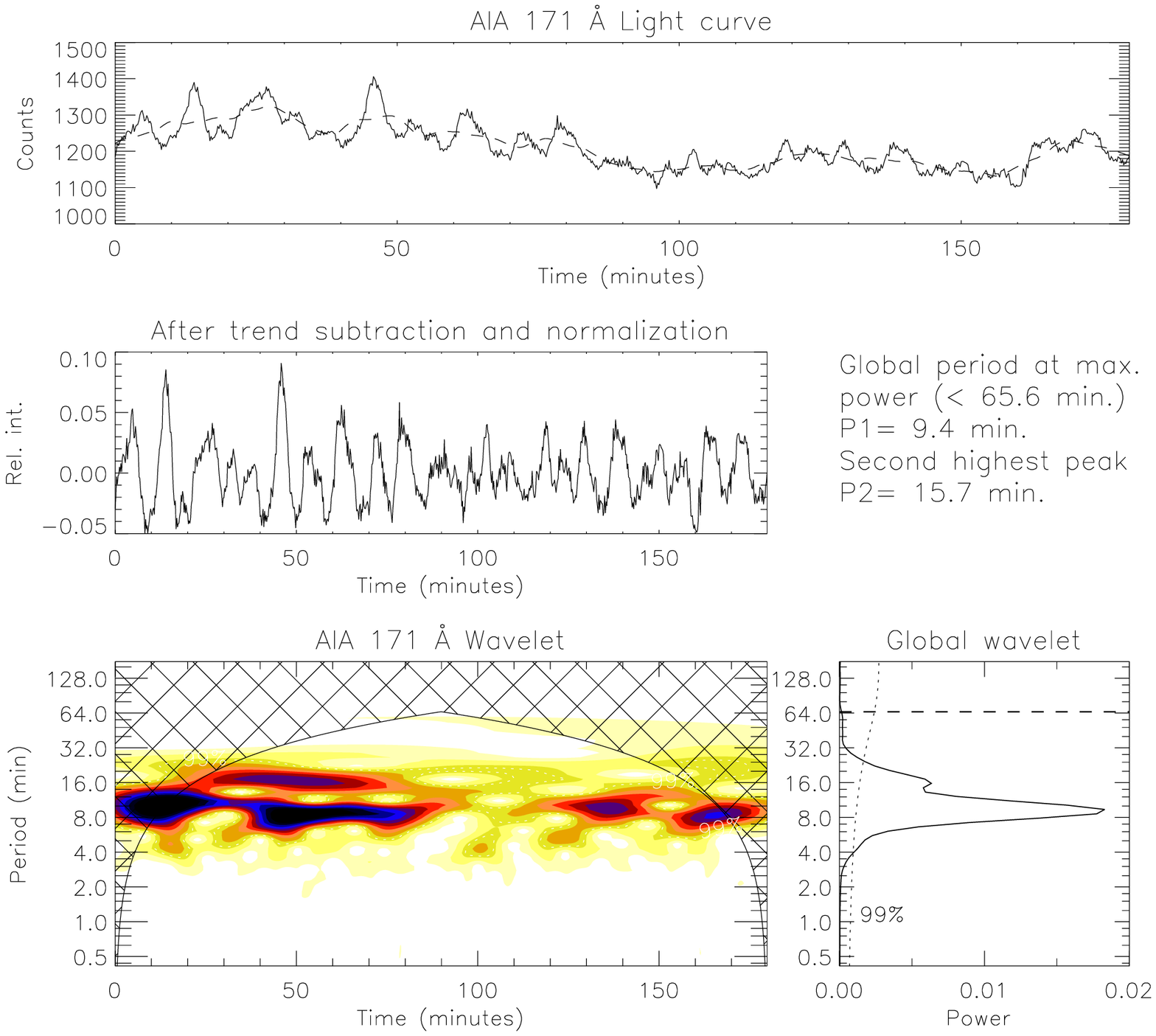}
\includegraphics[angle=90,height=8.0cm]{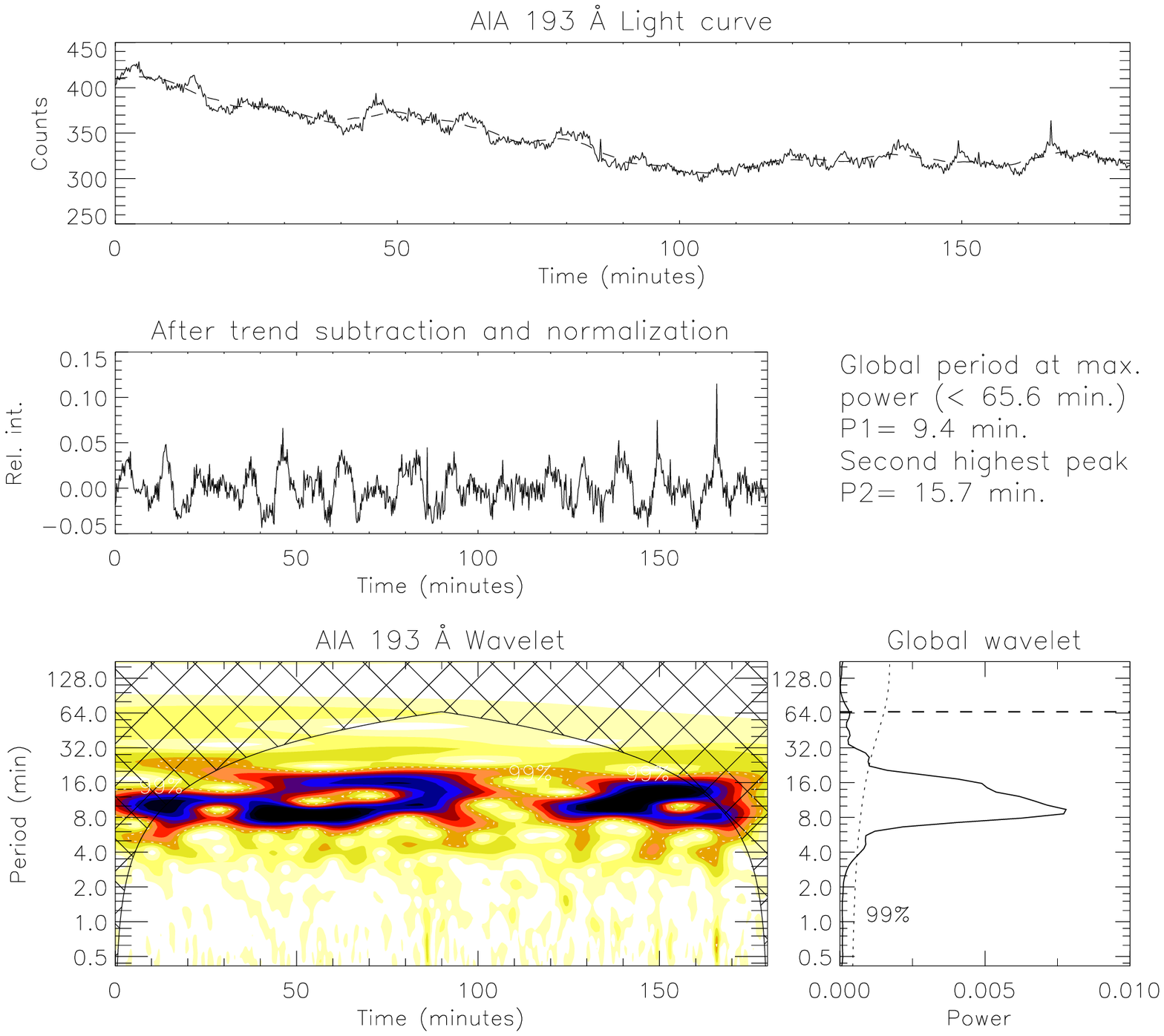}
\caption{Wavelet plots for the time series from the row marked by the dashed line in Fig~\ref{on-disk_plume} in 171~\AA\ (left) and 193~\AA\ channels. In each of these figures top panels show the original light curve in solid line and the subtracted trend in dashed line. The middle panels display the trend subtracted and normalized light curves. The main wavelet plot is shown in the bottom left and the global wavelet plot in the bottom right panels. Overplotted dotted line in the global wavelet plot is the 99\% significance curve. Top two strongest significant periods are listed as P1 and P2 in the middle right portion.}
\label{171_193_wavelet}
\end{figure*}
To estimate periodicity, we performed a wavelet analysis at an arbitrarily chosen row marked by a horizontal dashed line in the space-time maps. The wavelet plots at this location in the 171~\AA\ and 193~\AA\ channels are shown in Fig.~\ref{171_193_wavelet}. The first two significant periods obtained from this analysis are labelled at the middle right portion of these plots. These periodicity values are also listed in Table~\ref{values}. As can be seen from these values the periodicity of these disturbances is 9.4~min. The values listed in brackets in the table are the second-strongest peaks. Note the presence of the longer periods despite the 11~min filtering. They are only suppressed but not completely eliminated. 
\begin{figure}
\centering
\resizebox{\hsize}{!}{\includegraphics[angle=90, clip=true]{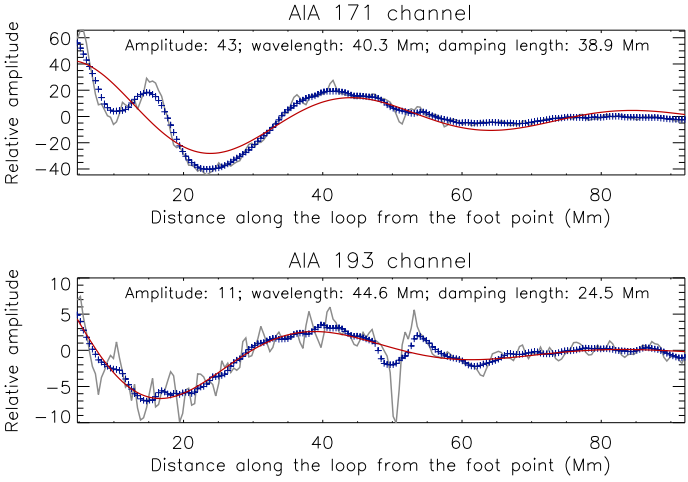}}
\caption{{\it Top:} Solid line in grey represents the intensity values along the loop, relative to the background, averaged over three successive time frames in 171~\AA\ channel. Plus symbols in blue are the 10 point ($\approx$ 4.3~Mm) smooth averages of these values. The damping sine curve fitted to these smoothed values is overplotted as a red solid line. Obtained fit parameters are labelled. {\it Bottom:} Similar curves and values generated from the 193~\AA\ channel.}
\label{damping}
\end{figure}

We also investigated the spatial damping along the loop. Fig.~\ref{damping} shows the intensity values along the loop (solid grey line) relative to the background, averaged over three successive time frames. Overplotted plus symbols in blue are the 10 point ($\approx$ 4.3~Mm) smooth average values. A damping sine wave function given by 
\begin{equation}
	I(l)=A_{0}\mathrm{sin}(\omega l+\phi)\mathrm{exp}(- l/L_{d}) + B_{0}+B_{1}l 
 \label{eq:function}
\end{equation}
was fitted to these values, which is also overplotted as a solid line in red. Here $A_{0}$ is the amplitude of the sine wave, $l$ is the length along the loop, $\omega$ and $\phi$ are frequency and phase of the wave, and $L_{d}$ represents the damping length. The last two terms were used to fit the background trend, if present. The first-order fit is found to be more than sufficient for this purpose. The best fits were obtained using a Levenberg-Marquardt least-squares minimisation \citep{1969drea.book.....B}. Top and bottom panels of Fig.~\ref{damping} display the plots from the 171~\AA\ and 193~\AA\ channels, respectively. Obtained fit parameters are labelled in the respective panels. Clearly, the relative amplitude of the oscillation and the damping length ($L_{d}$ value in equation~\ref{eq:function}) are lower and the wavelength is slightly longer in the hotter 193~\AA\ channel. The damping lengths are 38.9~Mm and 24.5~Mm in the 171~\AA\ and 193~\AA\ channels respectively. Lower damping lengths indicate faster damping in the 193~\AA~channel.
\section{Slow-wave model}
\label{S-model}
The farther propagation of the lower frequencies and the stronger dissipation in hotter channels may be understood as an effect of the damping of slow magneto-sonic waves by thermal conduction. We consider a simple 1D model \citep{demoortel2003,owen2009}, where radial structuring has been neglected. More complicated models have been studied by \citet{soler2008}. The dispersion relation of slow waves in such a simple configuration is given by 
\begin{equation}
	\omega^3-\imath \gamma d c^2_\mathrm{s} k^2 \omega^2 - \omega k^2 c^2_\mathrm{s} + \imath d k^4 c^4_\mathrm{s}=0,
\end{equation}
where $\omega$ is the angular wave frequency, $k$ is the wave number, $\gamma$ is the ratio of specific heats and $c_\mathrm{s}$ is the sound speed. The parameter $d=\frac{(\gamma -1)\kappa_\parallel T_0}{\gamma c^2_\mathrm{s} p_0}$ expresses the importance of the thermal conduction. It has the dimension of time. The quantities $p_0$ and $T_0$ are the background pressure and temperature, respectively. $\kappa_\parallel=\kappa_0 T_0^{5/2}$ is the thermal conduction coefficient. Low $d$ means a weak thermal conduction, high $d$ corresponds to strong thermal conduction. Since the observed waves are presumably driven from near the foot points of the loop, we considered the frequency $\omega$ to be fixed by the driver, and solved the dispersion equation for the wave number $k$. For the vanishing thermal conduction ($d\omega = 0$), we find $k^2=\omega^2/ c^2_\mathrm{s}$, which corresponds to the sound wave solution. In the limit of weak thermal conduction ($d\omega \ll 1$), the solution may be found as
\begin{equation}
	k=\frac{\omega}{c_\mathrm{s}} \left( 1-\imath\frac{d\omega}{2}(\gamma-1)\right)=\frac{\omega}{c_\mathrm{s}}-\imath \frac{1}{L_\mathrm{cond}}, \label{eq:damping}
\end{equation}
where we defined the thermal conduction damping length $L_\mathrm{cond}$. This formula is only marginally applicable in the solar corona, since the thermal conduction in coronal loops is quite strong \citep{vd2011}. For instance, the results from \citet{vd2011} indicate $d\omega \approx 0.7$ which is less than 1 but not very low. However, a lot of the physics relevant for these observations may be understood from this asymptotic formula. 

We first consider the effect of temperature. When the temperature increases (i.e. when we observe the structure in a higher temperature passband), the thermal conduction coefficient $\kappa_\parallel$ increases as well. This results in a higher value for $d$, and thus a lower value for the damping length $L_\mathrm{cond}$ in the limiting case of weak thermal conduction ($d\omega \ll 1$) that we currently consider. As found in the observations, the slow waves are more heavily damped in higher temperature passbands. Now, we consider the frequency dependence of the damping. In Eq.~\ref{eq:damping}, we have defined $L_\mathrm{cond}=2c_\mathrm{s}/d\omega^2(\gamma -1)$, where it is now obvious that the damping length is inversely proportional to the frequency squared, or directly proportional to the period squared. Indeed, in the observations we find that the longer periods propagate much higher (longer damping lengths). This simple model explains the qualitative features of the observations quite well. This model could be tested by measuring the damping lengths for different frequency bands, and showing that a power law with slope 2 is found. In principle, it also possible to use this simple model to measure the thermal conduction. By measuring the damping length in a loop for several frequency bands, the thermal conduction may be found as the intercept of the power law between the frequency and damping lengths.

\section{Conclusions}
\label{S-conclusions}
Open-loop or plume-like structures both off-limb and on-disk show that propagating disturbances with longer periods travel farther distances in the powermaps constructed in different periodicity ranges. The apparent speeds obtained for the propagating disturbance in the on-disk plume structure are higher in the hotter (193~\AA) channel. The amplitude of the disturbance is relatively low in this channel. The disturbance is also observed to be damped along the loop. The damping lengths were found to be 38.9~Mm and 24.5~Mm in 171~\AA\ and 193~\AA\ channels, respectively. The faster damping of the disturbance in hotter channels might indicate possible damping due to thermal conduction and matches the earlier theories on slow-mode wave damping. All these observed properties can be explained using a simple slow-wave model. We conclude that these disturbances are more likely to be signatures of slow mode waves than quasi-periodic high-speed upflows. Furthermore, these intensity waves are omnipresent at longer periods, higher up even in active regions. 

\begin{acknowledgements}
We thank the referee for his/her valuable comments, which have enabled us to improve the quality of the paper. The AIA data used here are courtesy of SDO (NASA) and AIA consortium. DB wishes to thank the KULeuven research council for a senior research fellowship (SF/11/005).
\end{acknowledgements}

\bibliographystyle{aa}
\bibliography{aa19885ref}

\end{document}